\documentclass{roajarticle}
\usepackage{graphicx}
\usepackage[section]{placeins}
\usepackage{natbib}
\usepackage{upgreek}
\newcommand{\volume}{\textbf{26}}
\newcommand{\numarul}{2}
\newcommand{\volyear}{2016}
\title{Dynamical analysis of NGC 110: Cluster of fainter stars
or data fluctuation?}
\newcommand{\shorttitle}{NGC 110}
\author[1,2]{Gireesh C. \MakeTextUppercase{Joshi}}

\affil[1] {P.P.S.V.M.I. College, Nanakmatta, U.S. Nagar, 262311, Uttarakhand, India\\
Email:gchandra.2012@rediffmail.com}
\affil[2] {H.N.B. Govt. P. G. College, Khatima, U.S. Nagar, 262308, Uttarakhand, India}


\begin{document}
\maketitle


\begin{abstract}
The stellar enhancement of the cluster $NGC~110$ {\bf is} investigated in various optical and infrared (IR) bands. The radial density profile of the IR region does not show a stellar enhancement in the central region of the cluster. This stellar deficiency may be occurring by undetected fainter stars due to the contamination effect of massive stars. Since, our analysis is not indicating the stellar enhancement below 16.5 mag of $I$ band, therefore the cluster is assumed to be a group of fainter stars. The proposed magnitude scatter factor would be an excellent tool to understand the characteristic of colour-scattering of stars. The most probable members {\bf do} not {\bf coincide} with the model isochronic fitting in the  optical bands due to poor data quality of $PPMXL$ catalogue. The different values of the mean proper motions are found for the fainter stars of the cluster and field regions, whereas similar values are obtained for radial zones of the cluster. The symmetrical distribution of fainter stars of the core are found around the best solution of isochrone. The mass function and mass segregation studies are not possible due to higher uncertainty of the photometric data. The number of the massive stars of the cluster region is low in comparison with the field region due to completed evolution life of the massive stars. 
\end{abstract}

\section{Introduction}

Open Star Clusters (OSCs) are excellent tools to understand the galactic evolutionary processes and dynamics of the stellar evolution. Furthermore, OSCs are dynamically associated groups of stars, which are bound to each other through the weak gravitation and located approximately at the same distance \citet{nil+2002}. The main sequence (MS) of {\bf clusters} is used for determining their parameters/information/facts. Each stellar sequence provides the evolutionary history of born stars from the same parent molecular cloud. Recently, \citet{jos+2015} found separate sequences of the cluster and field. They have correlated the mass function slope of DOLIDZE 14 with incompleteness of the data. \citet{Tadross+2011} and \citet{jos+2015a} estimated the basic physical parameters of the cluster NGC 110. They also separated the field stars from the cluster region by utilizing the combined effort of dynamical and statistical cleaning processes. Furthermore, they found stellar enhancement for extracted stars from the PPMXL catalogue \citep{Roeser+2010}. Although, the distance, age and reddening of studied cluster (NGC 110) are estimated through 2MASS \citep{Skrutskie+2006} data. The center coordinates (RA, DEC) of studied cluster NGC 110 are estimated by \citet{jos+2015a}, which comes as ($00{h}:27{m}:22.4{s}, +71^{o}:23^{'}:56.6^{''}$). Similarly, \citet{Tadross+2011} estimated the center of NGC 110 as $00{h}:27{m}:00{s},+71^{o}:23^{'}:00^{''}$. The radius of the cluster NGC 110 is obtained to be $\sim~10~arcmin$ by \citet{Tadross+2011}. The various parameters of NGC 110 are listed in the Table\ref{tab0}. A detailed membership analysis of each cluster becomes a robust investigative process to determine their properties \citep{car+2008}. \\
\begin{table}
\caption{The known physical parameters of the cluster NGC 110 under study. \label{tab0}}
\setlength{\tabcolsep}{1pt}
\small
\begin{center}
\begin{tabular}{lll}
\hline
Parameters & ~~~~\citet{jos+2015a}~~~~ \\
\hline
RA (J2000) & $00^{h}:27^{m}:22.4^{s}$ \\
DEC (J2000) & $+71^{o}:23^{'}:56.6^{''}$ \\
Core radius (arcmin) & $0.79{\pm}0.19$ \\
Cluster radius (arcmin)  & $5.6{\pm}0.4$ \\
Mean proper-motion (mas/yr) & $4.03{\pm}0.29$, $2.53{\pm}0.23$ ~~~~\\
Log(age) (yr) & $9.0{\pm}0.2$ \\
E(B-V) (mag) & $0.42{\pm}0.03$ \\
Distance-modulus (mag) & $10.57{\pm}0.30$  \\
Distance (kpc) & $1.29{\pm}0.22$ \\
\hline
\end{tabular}
\end{center}
\end{table}
The present study has been carried out through PPMXL data. PPMXL is a catalog of positions, proper motions, 2MASS and optical photometry of 900 millions stars and galaxies, aiming to be complete down to about V=20 mag full sky\footnote{irsa.ipac.caltech.edu/../ppmxl.html}. It is the result of a re-reduction of USNO-B1 together with 2MASS to the ICRS. In this catalogue, the stellar magnitudes are listed in the optical (B, R and I-bands) and near-infrared (J, H and $K_{s}$ bands) photometry. The outlines of the present paper are given as following. The nature of stellar enhancement is discussed in Section~\ref{sect:rad}. In Section~\ref{stat}, we have been iterated the statistically cleaning procedure for MPMs ({\bf i.e. most probable members}). The detailed proper motion analysis is carried out in Section~\ref{prop}. An analysis of the stellar distribution in the various radial zones are described in Section~\ref{stel}. The analysis of luminosity function and summary of results are prescribed in Section~\ref{lum} $\&$ \ref{conc}, respectively.
\section{Radius in Various bands}
\label{sect:rad}
\begin{figure}
\begin{center}
\includegraphics[width=15.0cm, angle=0]{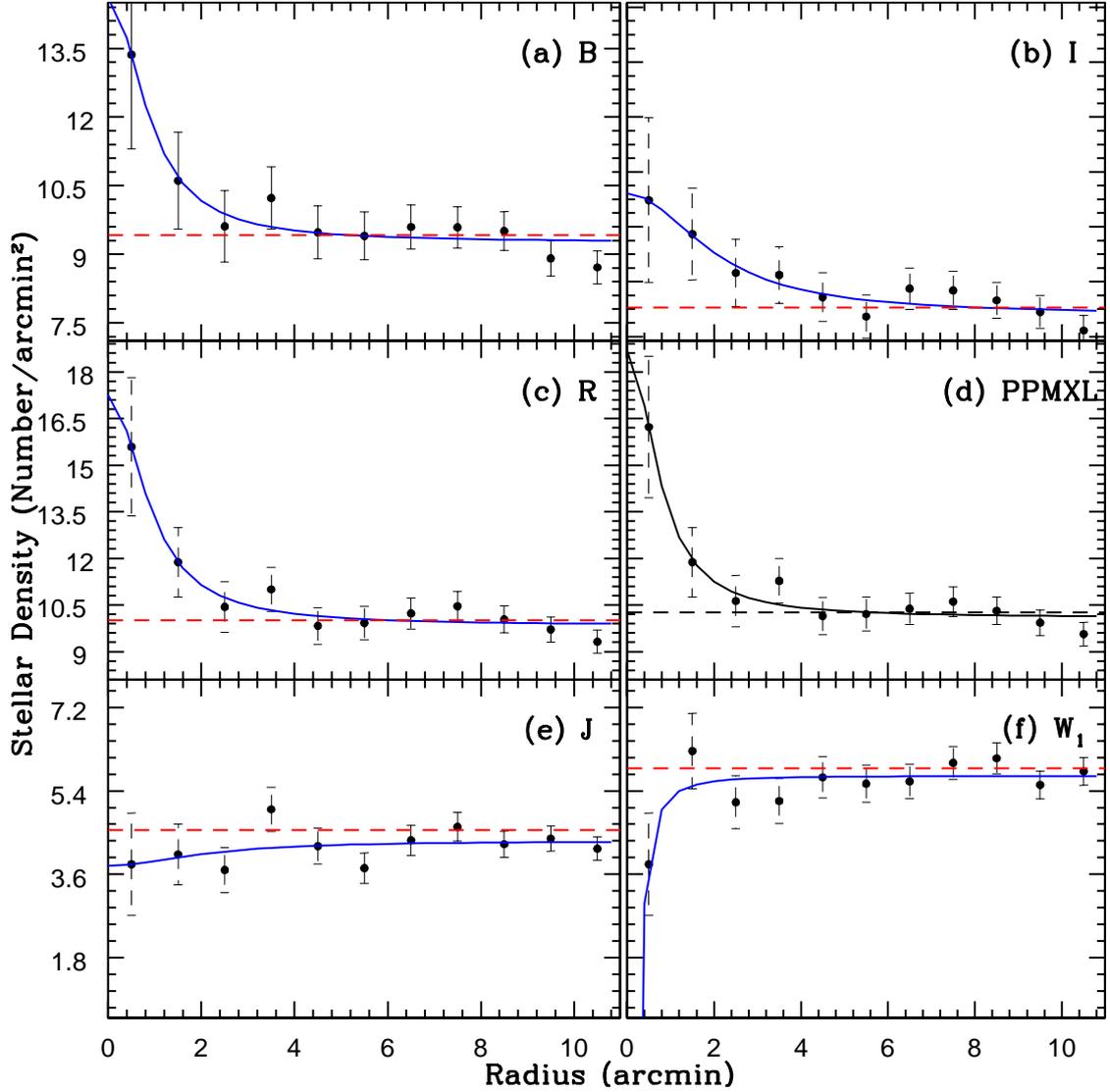}
   \caption{ The radial density profiles of the cluster region in the various photometric pass-bands. The solid curve in each panel represents the best fitted solution of the King-Empirical model on the computed radial stellar densities. {\bf Similarly, the red dashed lines mark the background field star density including 1-$\sigma$ error.}} 
   \label{Fig1}
\end{center}
   \end{figure}

The finding chart of this cluster is given in the manuscript of \citet{jos+2015a}. We have fitted the King-Empirical Model \citep{King+1966} on the radial density profile (RDP) for determining the radius of cluster. We have determined the radial density points for the radial zones of the studied cluster in each listed band of $PPMXL$. Moreover, these data points are used to construct the RDPs of the cluster in each individual band of the $PPMXL$ catalogue. As a result, the radius of cluster is varying according to the nature of RDPs as depicted in the Figure~\ref{Fig1}. For constructing the RDPs, the width of each radial zone is taken to be 1 arcmin and the center of studied cluster is extracted from the work of \citet{jos+2015a}. The stellar enhancement is occurred for the optical bands (B, R and I) only, whereas central dip is occurred in the RDPs of bands of 2MASS and WISE [Wide-field Infrared Survey Explorer; \citet{wri10}]. Further, the WISE is operating in the mid-IR, {\bf being} a NASA Medium Class Explorer mission which conducted a digital imaging survey of the entire sky in the 3.4, 4.6, 12 and 22 $\mu$m. {\bf WISE} has produced a reliable Source Catalog containing accurate photometry and astrometry of more than 300 million stellar objects. The RDPs of B, R and I optical bands are depicted in the (a), (c) and (b) panels of the Figure \ref{Fig1}, respectively; and the resultant parameters through these RDPs are listed in the Table \ref{tab1}. {\bf In the} panel (d) of figure \ref{Fig1} is shown the RDP of \citet{jos+2015a}. Other hand, the RDPs of infrared bands are depicted in the panel (e) and (f) of the Figure {\bf \ref{Fig1}, in which stellar} enhancement is not found. Thus, the stellar enhancement of cluster is found for the shorter wavelength (stellar radiation in visual bands) i.e. radiation of higher energetic photon instead of photons of IR region.  Since, the cool stars/red-giant stars are emitting more energy in the visual wavelengths rather than infrared, therefore, the cluster NGC 110 may be a group of either fainter stars or white dwarfs (WDs).  The conclusion of clustering of WDs is difficult due to quite faint limits of WDs ($\sim$22mag in optical band), while PPMXL values in this faint range are often doubtful.\\
It can be clearly seen, around the bright Tycho star (TYC 4303-1643-1 of $NGC~110$), PPMXL listed quite a few FALSE ``stars", most of them with faint magnitudes and some with very high proper motion values. If these high proper motion stars are real, they should be very close and most probably bright ones. Thus, an hastily made statement/result may be possible due to non-reliable entries in PPMXL catalogue whereas, 2MASS based RDP shows no``central enhancement". One cannot exclude the possibility that for those relatively bright stars (up to the 2MASS J band limit), there can be no apparent central enhancement, especially for those NOT well mass-segregated stellar system. On this background, we have needed to verify the stellar enhancement in the optical bands (B, R and I) by removing fainter stars as well as stars having very-high proper motion values. For this purpose, we have selected those stars which are brighter than 16.5 mag. Furthermore, we have found still enhancement in the center of the present studied cluster.

\subsection{ Concentration parameters}
The concentration parameter ($c$) of any cluster is defined as the logarithm value of the ratio of limiting radius ($r_ {lim} $) of the cluster to its core radius \citep{pet+1975}. We have found positive values of $c$ for visual photometric bands ($B$, $R$ and $I$). For this purpose, the $r_ {lim} $ has been computed through the following relation \citep{buk+2011}, 
$$r_{lim}= {\bf r_{core}} \sqrt{\frac{f_{0}}{3{\sigma_{bg}}}-1},$$
where ${\bf f_{0}}$, $r_{core}$ and $\sigma_{bg}$ are {\bf the central density}, core-radius of the cluster and uncertainty in the estimation of background stellar density, respectively. These values are useful to understand the nature of average stellar enhancement of the cluster. We have obtained different values of $c$ for different photometric bands (given in Table~\ref{tab1}) and the value of ${\bf c}$ is related to the number of detected stars in these bands. 
\begin{table}
\begin{minipage}[]{170mm}
\caption[]{The concentration and limiting radius of the cluster in the various photometric bands. These prescribed values are obtained throgh the RDP of listed stars in various optical bands(B, R and I bands) of PPMXL catalogue and these profiles are depicted in the panels (a, b and c) of figure 1. \label{tab1}}\end{minipage}
\setlength{\tabcolsep}{1pt}
\begin{center}
 \begin{tabular}{|c|c|c|c|c|c|c|}
  \hline
 Photometric band & Radius & Central density & Error in background & Concentration & Limiting radius & Core radius \\
                  & (arcmin)& ($f_{0}$) &  density ($\sigma_{bg}$)       &     & (arcmin)        & (arcmin) \\\hline
B & $5.2{\pm}0.4$ & $5.355{\pm}0.913$ & 0.157 & 0.553 & 3.214 & 0.900${\pm}$0.252\\
R & $6.4{\pm}0.4$ & $7.424{\pm}0.949$ & 0.173 & 0.596 & 3.647 & 0.925${\pm}$0.196\\
I & $8.4{\pm}0.4$ & $2.588{\pm}0.330$ & 0.163 & 0.173 & 2.732 & 1.835${\pm}$0.404\\
 \hline
\end{tabular}
\end{center}
\end{table}
\section{Statistical Separation and its utilization}
\label{stat}
The mean proper motion values of various radial zones are obtained through Gaussian fit technique, these values seem to be close to each other. We have adopted statistical magnitude-colour distance approach \citep{jos+2015b} to identify new most probable members (MPMs) in the visual bands. In this procedure, we have taken two types of grid size for a field star in $(R-I)-I$ space. The said grid sizes are given as below:
\begin{enumerate}
\item {The colour-excess $(R-I)$ and magnitude $I$ values for a grid dimension of a field star are considered to be ${\pm}0.02~mag$ and ${\pm}0.10~mag$), respectively. This said field star is situated at the center of grid and known as reference point.}
\item {For second type grid, the said values of $(R-I)$ and $I$ are considered to be ${\pm}0.01~mag$, and ${\pm}0.05~mag$, respectively.}
\end{enumerate}
In this procedure, the area of cluster and field regions are taken to be equal. The remaining stars are further separated from the cluster through their proper motion. For it, we have been utilizing the list of dynamical members as provided by \citet{jos+2015a}.\\  
\begin{figure}
   \centering
\includegraphics[width=15.0cm, angle=0]{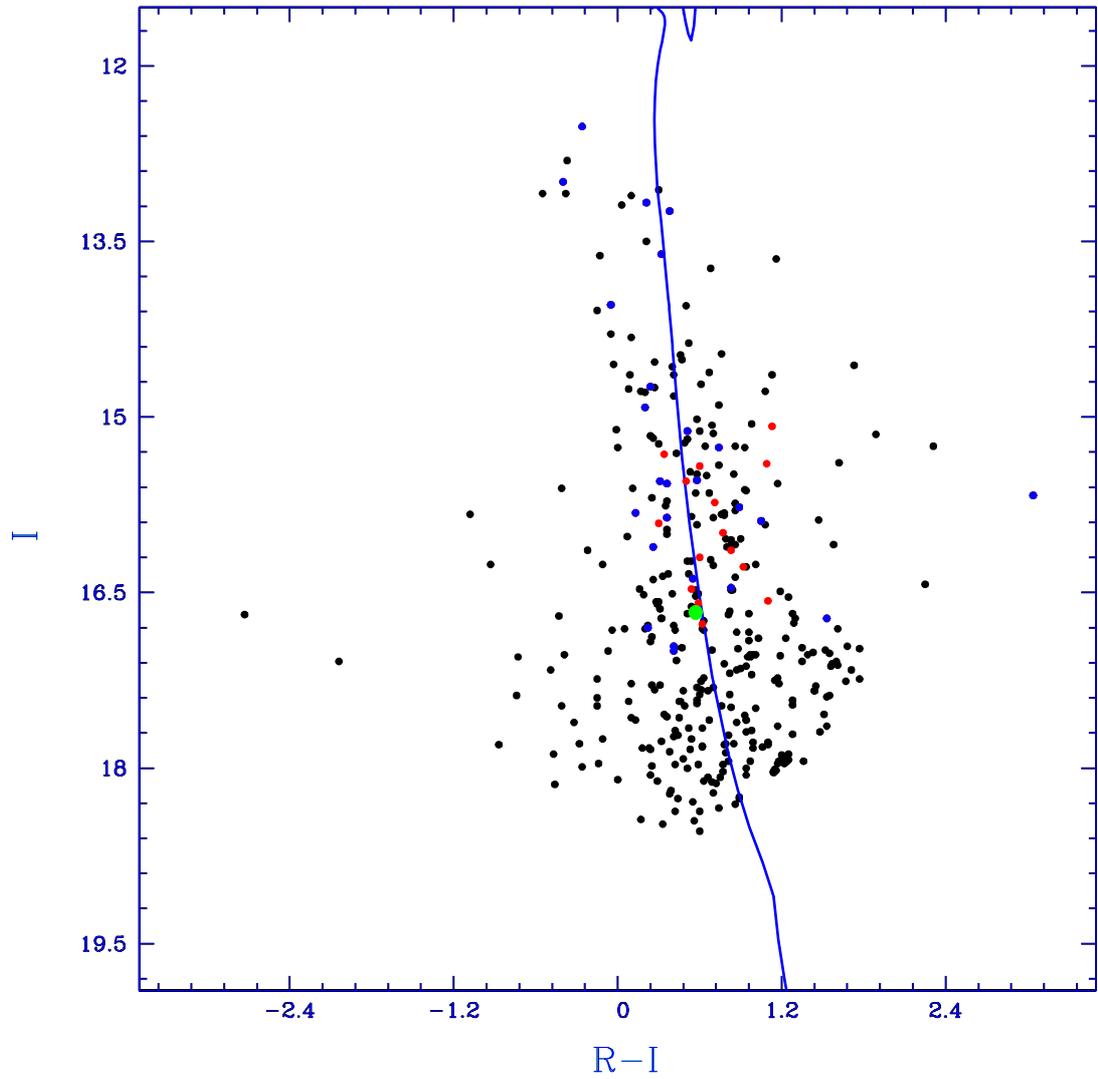}
   \caption{ The influence of the grid size in the identification of the most probable members of the cluster region. The blue solid curve represents the solution of the best fitted isochrone.} 
   \label{Fig2}
   \end{figure}
\begin{table}
\begin{minipage}[]{170mm}
\caption[]{The list of cross matched MPMs of Near infrared (NIR) photometry of the cluster {\citet{jos+2015a}} with the MPMs of present study. The later MPMs are obtained  by using the statistical cleaning process on the stellar magnitudes of R and I optical bands. \label{tab2}}\end{minipage}
\setlength{\tabcolsep}{1pt}
\begin{center}
 \begin{tabular}{|c|c|c|c|c|c|c|}
  \hline
 S. No. & RA  & DEC  &Proper Motion in RA (${\bf \mu{x}}$) & Proper Motion in DEC (${\bf \mu_{y}}$) & R & I\\
        & (Deg.) & (Deg.) & (mas/yr) & (mas/yr) & (mag)& (mag)\\\hline
01  &  6.963359  & 71.354309  &  0.8  &  5.1  &  12.24  &  10.46\\
02  &  6.798335  & 71.364241  &  4.6  &  3.0  &  12.25  &  11.43\\
03  &  6.855886  & 71.390639  &  0.7  &  1.8  &  09.98  &  09.93\\
04  &  6.800651  & 71.456568  &  4.4  &  4.2  &  12.26  &  12.52\\
05  &  6.847568  & 71.424183  & -0.1  &  2.0  &  13.38  &  13.17\\
06  &  6.895260  & 71.374188  & -1.9  &  2.0  &  13.62  &  13.24\\
07  &  6.785487  & 71.398294  &  0.5  &  5.6  &  12.59  &  12.99\\
08  &  6.880143  & 71.385233  & -2.1  & -2.8  &  13.93  &  13.61\\
09  &  6.830988  & 71.431911  & -0.3  &  1.7  &  13.99  &  14.04\\
10  &  6.838362  &  71.43464  &  6.0  &  2.8  &  14.98  &  14.74\\
11  &  6.772891  & 71.389527  &  1.7  &  1.1  &  15.63  &  15.12\\
12  &  6.716714  & 71.439326  &  4.4  & -1.2  &  15.12  &  14.92\\
13  &  6.960778  & 71.371951  &  9.4  &  1.9  &  16.00  &  15.26\\
14  &  6.814866  & 71.462263  &  6.4  &  5.1  &  16.12  &  15.54\\
15  &  6.853756  & 71.340099  &  5.0  &  5.7  &  18.71  &  15.67\\
16  &  6.695908  & 71.426849  &  1.6  &  3.7  &  15.95  &  15.82\\
17  &  6.711744  &  71.37339  &  3.4  &  5.3  &  15.93  &  15.57\\
18  &  6.698227  & 71.421794  &  4.5  &  1.5  &  16.22  &  15.86\\
19  &  6.758425  & 71.340613  &  1.5  &  4.4  &  15.86  &  15.55\\
20  &  6.940687  & 71.338757  &  3.4  &  2.6  &  16.66  &  15.77\\
21  &  7.021364  & 71.395864  &  1.7  &  5.5  &  16.94  &  15.89\\
22  &  6.831960  & 71.358905  &  5.9  & -0.7  &  18.25  &  16.72\\
23  &  6.654649  & 71.419753  &  4.3  &  0.1  &  16.37  &  16.11\\
24  &  6.771381  & 71.357042  &  5.6  &  7.3  &  17.29  &  16.46\\
25  &  6.773691  & 71.353251  &  2.7  &  2.9  &  16.93  &  16.38\\
26  &  6.775755  & 71.366764  &  4.5  &  3.9  &  17.41  &  17.00\\
27  &  6.765779  & 71.397239  &  4.3  &  8.1  &  17.37  &  16.96\\
28  &  6.802226  & 71.397738  &  7.7  &  2.8  &  17.02  &  16.80\\
29  &  6.841189  & 71.429891  &  8.2  &  1.6  &  17.24  &  16.67\\
30  &  6.878371  & 71.419160  &  5.3  &  3.5  &  16.02  &  15.42\\
31  &  6.753539  & 71.377299  &  6.8  & -2.0  &  16.49  &  15.40\\
32  &  6.776112  & 71.406968  &  0.4  &  6.2  &  15.66  &  15.32\\
33  &  6.785655  & 71.390304  &  3.1  &  3.8  &  16.44  &  15.73\\
34  &  6.855808  & 71.429026  &  3.3  &  4.1  &  16.05  &  15.55\\
35  &  6.910567  & 71.432888  &  7.8  &  3.7  &  16.21  &  15.91\\
36  &  6.931771  & 71.421137  &  4.7  &  4.1  &  17.67  &  16.57\\
37  &  6.725314  & 71.364254  &  0.9  & -0.8  &  16.21  &  15.08\\
38  &  6.876448  & 71.355959  &  4.0  &  4.5  &  16.76  &  15.99\\
39  &  6.873529  & 71.433641  &  5.4  & -3.6  &  17.18  &  16.59\\
40  &  6.732527  & 71.370386  &  3.7  &  1.8  &  16.97  &  16.14\\
41  &  6.995426  & 71.414484  &  6.0  &  4.8  &  17.20  &  16.28\\
42  &  6.887853  & 71.433961  &  0.7  &  2.4  &  16.80  &  16.20\\
43  &  6.652871  &  71.38503  &  3.3  &  6.0  &  17.39  &  16.77\\
44  &  6.892048  &  71.38624  &  3.7  &  2.7  &  17.01  &  16.47\\
    \hline
\end{tabular}
\end{center}
\end{table}
We have compared new MPMs with the list of MPMs through near-IR (NIR) photometry. We have found the scattered distribution of those MPMs in $(R-I) -I$ CMD which shows the closest trend with isochrone fit in the NIR region. Furthermore, the cross matched MPMs are decreasing with the increment of the grid size. We have found {\bf 44} common MPMs (all stars) of Figure~\ref{Fig2} except black and green dots) for the case of former grid size, whereas there are only 29 stars (shown by red and green dots in Figure~\ref{Fig2}) for the latter grid size. Our investigation indicates that there are 28 common stars (stars having S. No., 1 to 28 in Table~\ref{tab2}) between both lists of MPMs obtained through different grid sizes. It is interesting fact that 29$^ {th} $ star of Table~\ref{tab2} is found as a MPM for big grid size and not for smaller grid size and it is depicted by the big green dot in Figure~\ref{Fig2}. Such typical facts may be arising due to the poor photometric quality of $PPMXL$ catalogue.\\
After cross matching of stars between $R$ and $I$ bands within 1 arcsec scatter, a total of 806 and 800 stars are found in the cluster and field regions respectively. Since cluster and field regions are holding equivalent area, therefore, it is an unexpected result of stellar enhancement (~0.75 $\%$) of cluster compared to the field region, whereas, well stellar enhancement are seen in individual $R$ and $I$ bands. It seems that there are several stars which are only visible in a particular band, which happened either due to detection of several stars in a particular wavelength range of radiation or or due to FALSE stars of PPMXL catalogue.
\subsection{Cluster characteristics versus detected stars}
\begin{figure}
   \centering
\includegraphics[width=15.0cm, angle=0]{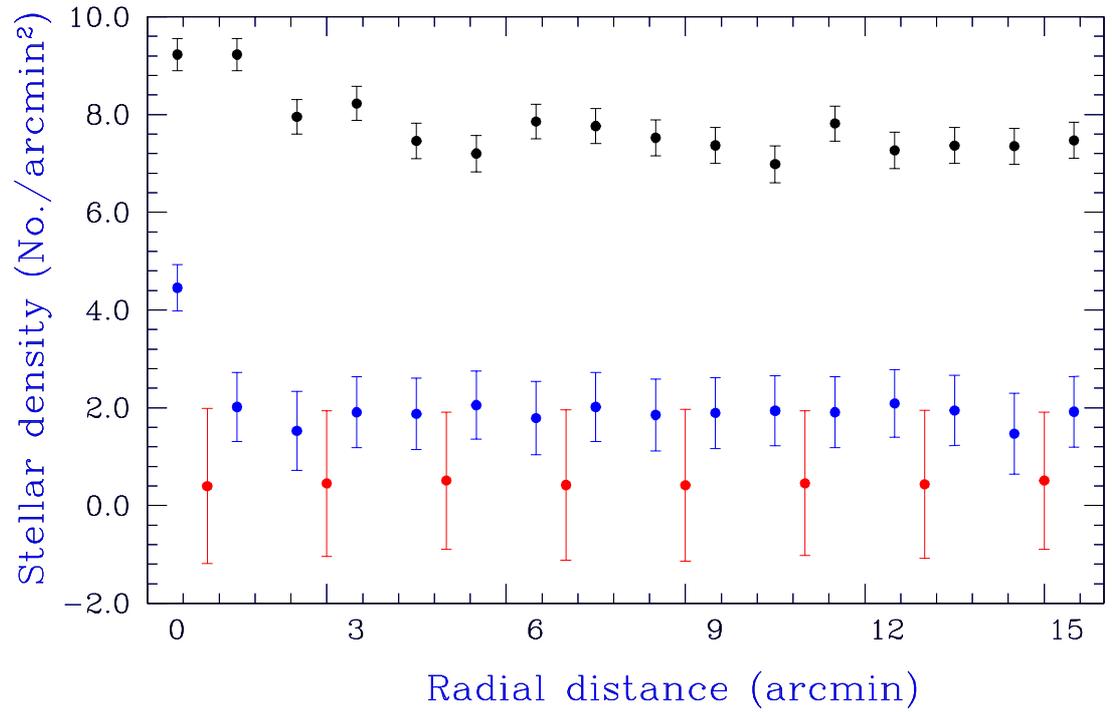}
\vspace{-5cm}
   \caption{ The radial density profiles of the stars. The blue, red and black dots represent the detected stars in only $R$, only $I$ and both photometric bands respectively.} 
   \label{Fig3}
\end{figure}

After cross matching of stars between $R$ and $I$ bands, we have divided them into three groups as follows (i) common stars in both bands, (ii) only detected in $R$ band and (iii) only detected in $I$ band. We have drawn RDPs for all these groups and the smooth curves of RDPs are not obtained for these groups as shown in Figure~\ref{Fig3}. Though, the density peak of core is well obtained in the RDP of common stars with zigzag pattern. In addition, this zigzag pattern of common stars becomes the smooth curve after stellar doping from the other prescribed groups. Since, FALSE stars are also presented in the PPMXL catalogue, therefore cautious is needed to interpret the PPMXL based {\bf on} RDPs. The RDP of common stars does not show the possibility of false stellar enhancement of the cluster due to detection of stars in multiple optical bands. The deep $UBVRI$ observations of this system are needed for more clear evidence.
\subsection{Magnitude scatter factor}
\label{msf}
It seems that the photometric magnitude of stars depends on the effective wavelength of pass bands. The co-relation analysis among the stellar magnitudes of various photometric bands may effectively explain the dependency of stellar magnitudes from one band to another  band. The fixed limits of various magnitude bins of a photometric band is taken to find out their stellar range in other photometric bands. This variation may be occurring either due to the precision error in estimation of stellar magnitude or due to the scattering effect (which is produced due to the presence of interstellar clouds/dust in the direction of field of view of the cluster). Furthermore, the range of these bins are either increasing or decreasing with the increasing magnitude bins.\\
We are proposed a new parameter ``magnitude scatter factor (MSF)" for understanding the data quality and effect of chemical composition of the environment of cluster. It is defined as the ratio of magnitude range of the variable bin of a photometric band to the magnitude range of the respective bin of a reference band. The $H$ - band of 2MASS is taken as a reference band and other bands are considered to be secondary bands (having variable range of bins). For the determination of ``MSF'', we are divided stars into various magnitude bins through extracted B-band data of PPMXL. The stars of prescribed bins are cross matched with the stellar magnitudes of other photometric bands. Moreover, these cross-matched stars are arranged in the descending/ascending order of magnitudes of a particular band. The difference of highest and lower magnitude of these arranged stars, is defined as the $dX$ (where $X$ stands for any band from $R,~I,~J,~H,~K$). The obtained MSF values of these bands are listed in Table~\ref{tab3}. It is observed that the MSF values are increasing for shorter wavelength ($R$, $I$, $J$ etc.) but decreasing for longer wavelength ($K$ etc.). Since, the detection rate of stars are decreased {\bf towards the fainter} magnitudes, therefore the values of MSF may also be correlated with the incompleteness of data.\\
\begin{table}
\begin{minipage}[]{170mm}
\caption[]{The magnitude scatter factors (MSF) for the various photometric bands in the reference to $H$ band. \label{tab3}}\end{minipage}\\
\setlength{\tabcolsep}{2pt}
 \begin{tabular}{|c|c|c|c|c|c|c|c|c|c|c|c|}
  \hline
$B$-magnitude Range & $dR$ & $dI$ & $dJ$ & $dH$ & $dK$ & $\frac{dR}{dH}$ & $\frac{dI}{dH}$ & $\frac{dJ}{dH}$ & $\frac{dK}{dH}$ \\\hline
08-11 & 3.640 & 3.310 & 2.190 & 2.452 & 2.513 & 1.484 & 1.349 & 0.893 & 1.025 \\
11-13 & 2.890 & 1.910 & 1.907 & 1.957 & 2.109 & 1.477 & 0.976 & 0.974 & 1.077 \\
13-15 & 4.190 & 3.420 & 2.262 & 1.930 & 1.883 & 2.171 & 1.770 & 1.172 & 0.975 \\
15-17 & 3.580 & 2.625 & 1.437 & 1.424 & 1.089 & 2.514 & 1.843 & 1.009 & 0.765 \\
 \hline
\end{tabular}
\end{table}
These values of shorter wavelengths are increasing from brighter stars to fainter stars, whereas vice-verse for longer wavelengths. The stellar enhancement are occurred for those bands in which the MSF values are increasing with the stellar magnitudes. The bin size depends on the crowding strength of members, which may be highly influenced by the estimation precision of stellar magnitude in various photometric bands. Furthermore, the bin size of secondary bands are overlapped to each other for fixed and separate bins of the reference band.
\section{Proper Motion Study}
\label{prop}
\begin{figure}
\begin{center}
\includegraphics[width=10.0cm, angle=-90]{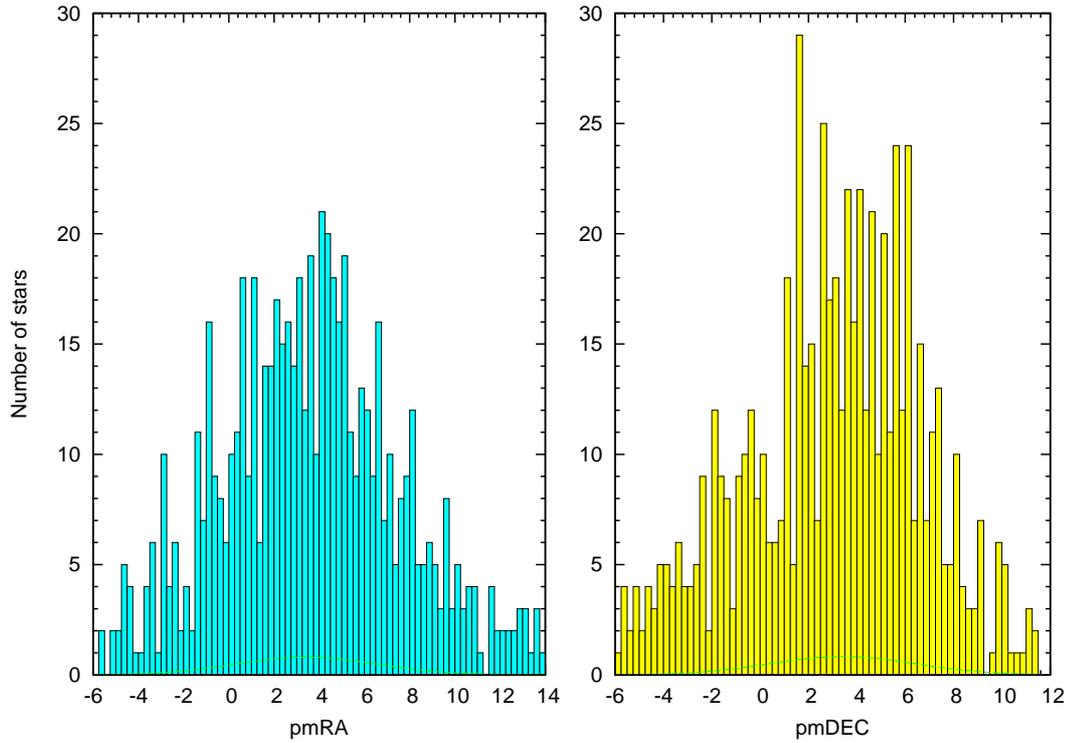}
   \caption{{\bf The histograms} for the pmRA (proper motion in RA) and pmDEC (proper motion in DEC).} 
   \label{Fig4}
\end{center}
\end{figure}

The angular displacement of stars per year is defined as the proper motion of stars. The proper motion of any star is divided into two components according to their motion in Right-Ascension (RA) and Declination (DEC) directions. The mean proper motion of this cluster is estimated by \citet{jos+2015a} through the mean iteration method. Here, we are estimated the mean proper motion through Gaussian method (shown in the Figure~\ref{Fig4}). The histograms of figure 4 is constructed through the proper motion values extracted from the PPMXL [\cite{Roeser+2010}] catalogue. The mean proper motions are found to be $3.375{\pm}0.195~{\bf mas/yr}$ and $3.493{\pm}0.178~{\bf mas/yr}$ in the RA and DEC directions respectively; which are deviated from the given values of \citet{jos+2015a}. The proper motion center of the cluster shifts toward to the center of the field stars. These values are also found to be $3.418{\pm}0.121~ mas/yr$ and $3.197{\pm}0.131~mas/yr$ through iteration method. Thus, these different results are found due to the addition of fainter stars instead of different procedure. These prescribed stars are not detected in the photometric bands of 2MASS.\\
\begin{figure}
\begin{center}
\includegraphics[width=15.0cm, angle=0]{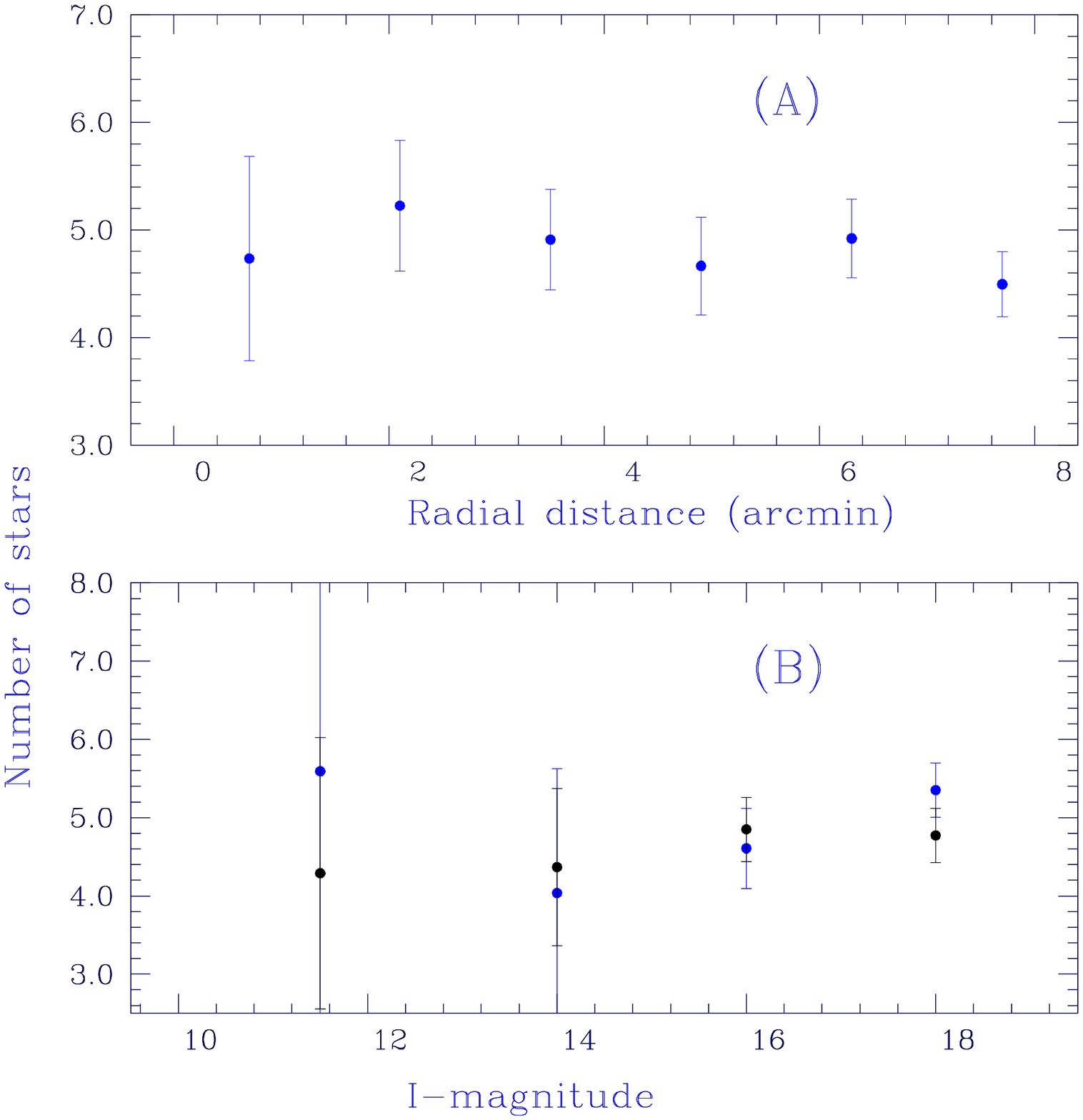}
   \caption{The variation of mean proper motion values with respect to (A) radial zones and (B) the magnitude bins respectively. In Figure 5(B), the cluster region and field region points are depicted by the black and blue points, respectively.}
   \label{Fig5}
\end{center}
\end{figure}
To understand the effect of field stars, we have estimated the mean proper motion of stars for various concentric radial zones as well as various magnitude bins as follows.
\subsection{Change with radial zones}
The field stars fraction is continuously increasing from the cluster center to the periphery. The field stars are symmetrically distributed in the region of cluster, whereas their {\bf fraction varies} in radial zones due to the exponential distribution of members of cluster. The mean-proper motions values of radial zones have represented in the Figure~\ref{Fig5}-A and listed in the Table~\ref{tab4}. The mean proper motions of field stars seems to be similar to the stars of the cluster. Since, the members of the cluster must {\bf have} different mean proper motions compared to the field stars, therefore, the whole cluster region seems to be highly influenced by the field stars. This result seems to be indicator of stellar enhancement due to background stellar fluctuation.\\
\begin{table}
\begin{minipage}[]{170mm}
\caption[]{The mean proper motion values of the stars of the cluster region in the various radial zones. \label{tab4}}\end{minipage}
\setlength{\tabcolsep}{1pt}
 \begin{tabular}{|c|c|c|c|c|}
  \hline
 Radial Range & Proper Motion in RA & Proper Motion in DEC & Resultant Proper Motion \\
 (arcmin) & ($\mu_{x}$ in mas/yr) & ($\mu_{y}$ in mas/yr) & ($\mu$ in mas/yr)\\\hline
0.0-1.4 & 3.208${\pm}$0.711 & 3.482${\pm}$0.632 & 4.735${\pm}$0.951\\
1.4-2.8 & 3.452${\pm}$0.439 & 3.922${\pm}$0.421 & 5.225${\pm}$0.608\\
2.8-4.2 & 3.629${\pm}$0.325 & 3.308${\pm}$0.335 & 4.910${\pm}$0.467\\
4.2-5.6 & 3.229${\pm}$0.351 & 3.368${\pm}$0.288 & 4.666${\pm}$0.454\\
5.6-7.0 & 3.474${\pm}$0.260 & 3.485${\pm}$0.254 & 4.925${\pm}$0.363\\
7.0-8.4 & 3.249${\pm}$0.231 & 3.106${\pm}$0.194 & 4.495${\pm}$0.302\\
 \hline
\end{tabular}
\end{table}
\citet{jos+2015a} obtained different mean proper motions of probable members compare to the field stars. Since, some new fainter stars are added in the list of MPM of the cluster due to their detection in $BRI$ photometry only and leading to the stellar enhancement. These new stars are not kinetically distinguished into field stars and members due to the similar proper motions of stars of both (cluster and field) regions. 
\subsection{Change with magnitude bins}
The mean proper motion of members of the various magnitude bins is very useful to know {\bf for} the dynamical behavior of massive and lighter stars. We have estimated the mean proper motions for various magnitude bins of both cluster and field stars. The mean proper motions of massive stars are found to be uncertain for both regions due to their low numbers, whereas these values seems similar for those stars having $I-$ magnitude between 13 to 17 mag as depicted in Figure~\ref{Fig5}-B and listed in Table~\ref{tab5}. Moreover, {\bf with} blue and black points of Figure~\ref{Fig5}-B are depicted the cluster region and field region points, respectively. The mean proper motion values of fainter stars are different for both regions. Since, fainter members make stellar enhancement and highly contaminated by the field stars. The uncertainty in the estimation of stellar magnitude (especially $BRI$ magnitude of $PPMXL$ catalogue) of fainter stars is high, which may reduce the effectiveness of field stars decontamination through the statistical approach {\bf \citep{jos+2015b}} due to the smallest size of the grid around a field star.\\
\begin{table}
\begin{minipage}[]{170mm}
\caption[]{The mean proper motion values of the stars of the cluster and field regions in the various magnitude bins. \label{tab5}}\end{minipage}
\setlength{\tabcolsep}{1pt}
\\
{\bf For the cluster region}\\
 \begin{tabular}{|c|c|c|c|c|}
  \hline
 I-magnitude & Proper Motion in RA & Proper Motion in DEC & Resultant Proper Motion \\
 (Mag) & ($\mu_{x}$ in mas/yr) & ($\mu_{y}$ in mas/yr) & ($\mu$ in mas/yr)\\\hline
   10-13 & -0.436${\pm}$3.097 & 5.574${\pm}$1.443 & 5.591${\pm}$3.417\\
   13-15 &  2.655${\pm}$1.204 & 3.042${\pm}$1.035 & 4.038${\pm}$1.588\\
   15-17 &  3.560${\pm}$0.418 & 2.925${\pm}$0.297 & 4.608${\pm}$0.513\\ 
   17-19 &  3.675${\pm}$0.257 & 3.891${\pm}$0.233 & 5.352${\pm}$0.347\\
 \hline
\end{tabular}\\
\\
\\
\setlength{\tabcolsep}{1pt}
{\bf For the field region}\\
 \begin{tabular}{|c|c|c|c|c|}
  \hline
 I-magnitude & Proper Motion in RA & Proper Motion in DEC & Resultant Proper Motion \\
 (Mag) & ($\mu_{x}$ in mas/yr) & ($\mu_{y}$ in mas/yr) & ($\mu$ in mas/yr)\\\hline
   10-13 & 2.888${\pm}$1.335 & 3.172${\pm}$1.107 & 4.290${\pm}$1.734\\
   13-15 & 2.695${\pm}$0.877 & 3.437${\pm}$0.493 & 4.368${\pm}$1.006\\
   15-17 & 3.878${\pm}$0.287 & 2.912${\pm}$0.294 & 4.850${\pm}$0.411\\
   17-19 & 3.198${\pm}$0.257 & 3.544${\pm}$0.233 & 4.774${\pm}$0.347\\
\hline
\end{tabular}
\end{table}
\section{Stellar Sequence in CMD}
\label{stel}
\begin{figure}
\begin{center}
\includegraphics[width=15.0cm, angle=0]{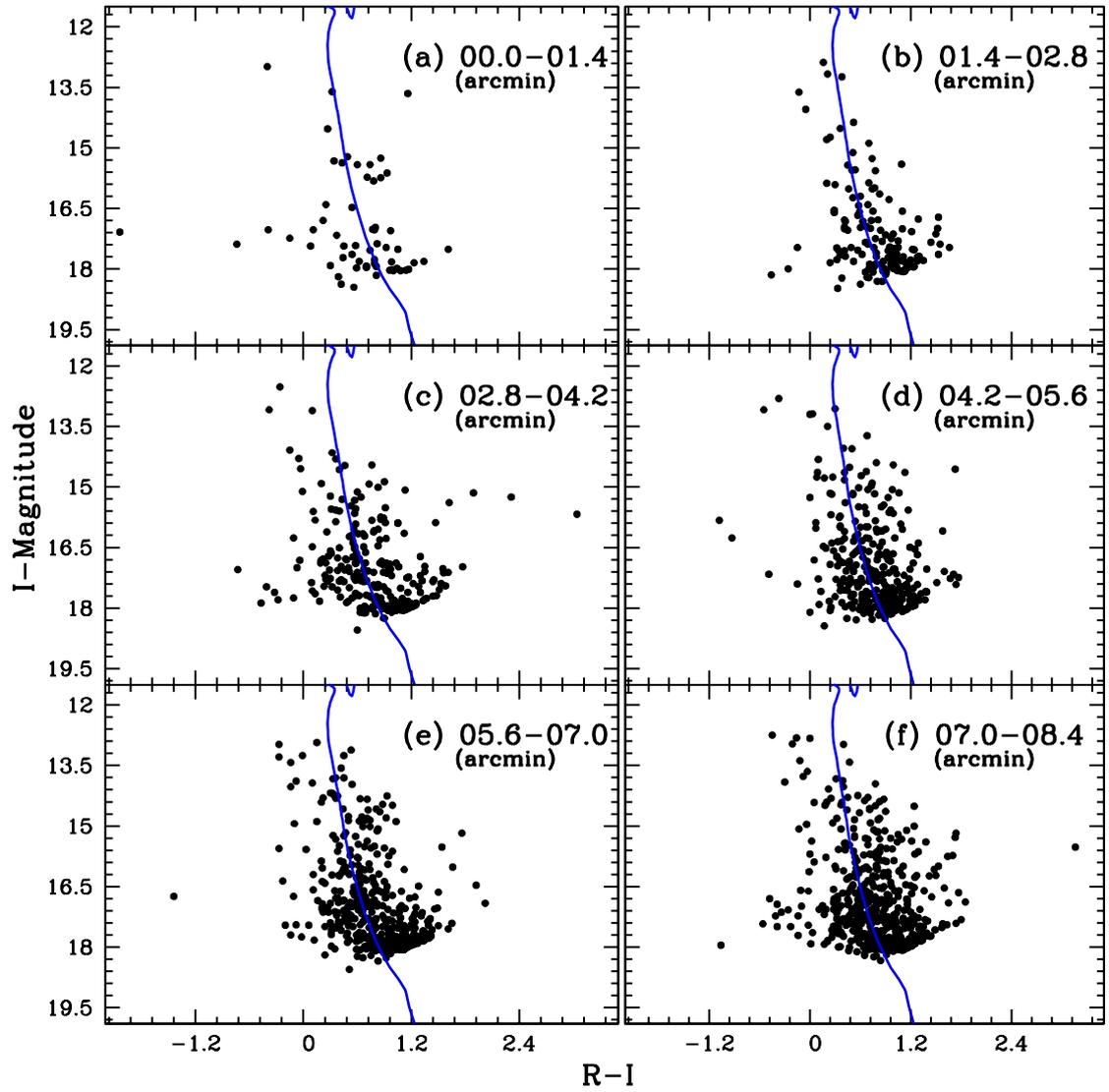}
   \caption{The appeared stellar sequences in various radial zones of the field view of the cluster region. The data of these diagrams is extracted from the $PPMXL$ catalogue.} 
   \label{Fig6}
\end{center}   
\end{figure}

Stellar sequence represents those stars which are born from the same molecular clouds at the same time but with different masses. The innermost radial zone of any cluster is dominated by its members. This sequence is least affected by the field stars. Field stars have two sequences as foreground (bluer) and background (redder) stars. The photometric broadening of CMD increases with the order of radial rings. The MS must be disappeared beyond the cluster extent. These sequences are visually separated to each other for massive stars except lighter stars. The stellar population is frequently increased for foreground sequence compared to background sequence as shown in the panels {\bf of} Figure~\ref{Fig6}. The various CMDs of Figure 6 are constructed through the $R$ and $I$ band data of the PPMXL catalogue. A total of 36, 150, 246, 311, 435 and 518 stars are found in the $(R-I)$ vs $I$ CMDs of radial zones $0.0-1.4$ arcmin, $1.4-2.8$ arcmin, $2.8-4.2$ arcmin, $4.2-5.6$ arcmin, $5.6-7.0$ arcmin and $7.0-8.4$ arcmin, respectively. Though, the total stellar numbers of these radial zones are 93, 195, 335, 437, 582 and 720, respectively. The less number of background stars of the cluster indicates the presence of stellar dust. {\bf This stellar dust (which absorbs the radiant photons of background stars) reduce the probability of detection of fainter background stars.} The dip in stellar density of the RDP with the above fact also provides a clue for existing higher density of the stellar dust within the cluster region. This inter-stellar dust may be so much prominent that the background field stars are not appearing in the CCD observations.\\
The MS stars are used to estimate the age, reddening and distance of the cluster and field star decontamination is carried out through the colour-magnitude-distance approach \citep{jos+2015b}. The distance and colour-excess values have been computed through the solution of the best fitted isochrone of solar mettalicity \citep{mar+2008} with log(age) of NGC 110 as 9.0 (equivalent to 1 Gyr) \citep{jos+2015a}. The apparent distance modulus and colour-excess i.e. ${E(R-I)}$ are found to be $10.95~mag$ and $0.13~mag$ respectively. The absolute distance modulus and cluster distance are computed through the \citet{jos+2015} relations and these values are found to be $10.612~mag$ and $1.325~kpc$ respectively. The computed distance shows close agreement with \citet{jos+2015a}. The very old age and distance of the cluster {\bf show a mixture} of fainter stars and remaining stellar dust.
\section{Luminosity Function}
\label{lum}
\begin{figure}
\begin{center}
\includegraphics[width=15.0cm, angle=0]{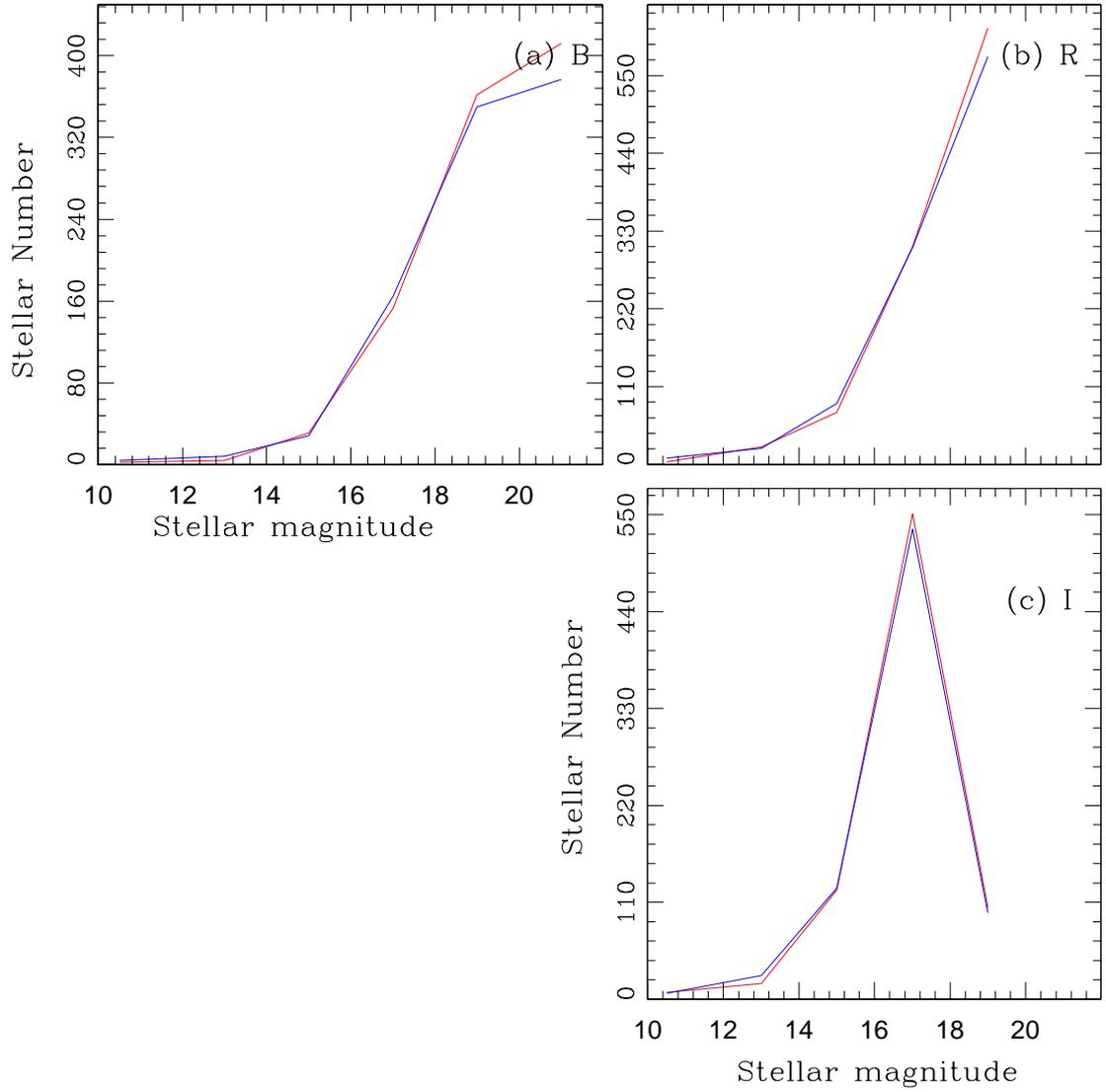}
   \caption{ The luminosity functions of the various magnitude bins of the various photometric bands. The red and blue solid lines are representing the LF for the cluster and field region respectively.} 
   \label{Fig7}
 \end{center}  
\end{figure}

The luminosity function (LF) is {\bf the number of cluster stars in different magnitude bins}. The LF of the cluster is high compare to the field, whereas this fact seems to be falsified for the cluster in infrared. The Figure~\ref{Fig7} shows high number of fainter stars (found in two last magnitude bins of CMD) of cluster {\bf in comparison with the} field region, whereas vice verso results {\bf occur} for massive stars. The obtained stellar number for the cluster and field region are listed in Table~\ref{tab6}. \\
\begin{table}
\begin{minipage}[]{170mm}
\caption[]{The values of LFs in the various photometric bands from the cluster region. The corresponding values for the field region are given in the parentheses. \label{tab6}}\end{minipage}
\setlength{\tabcolsep}{5pt}
 \begin{tabular}{|c|c|c|c|c|}
  \hline
Magnitude Bin & Stellar Number & Stellar Number & Stellar Number\\
(Mag)  & (B-band) & (R-band) & (I-band) \\\hline
09-12 & 002 (004) & 004 (009) & 008 (007) \\
12-14 & 004 (008) & 025 (023) & 018 (027) \\
14-16 & 031 (028) & 074 (086) & 124 (126) \\
16-18 & 153 (165) & 308 (307) & 551 (534) \\
18-20 & 362 (350) & 617 (577) & 104 (098) \\
20-22 & 412 (377) & --- & --- \\
 \hline
\end{tabular}
\end{table}
The detection of fainter stars of the cluster region decreases with the increasing effective wavelength of the photometric bands. By the comparison of LFs of $B$, $R$ and $I$ band, it is concluded that the fainter stars of the cluster region are showing continuously a disappearance towards the longer wavelength. Furthermore, this result also weights our conclusion that the dynamical processes of the stars of the cluster are occurring in such a way that most of the radiant energy of the cluster may be radiated below the effective wavelength of the IR range. The sudden and sharp decrements of the stellar number of the last magnitude bin of the LF of the $I$ photometric band may provide complementary result for our hypothesis.
\section{Discussion and Conclusion}
\label{conc}
It is well known fact that the proto-stars emit radiation in the IR range. The RDPs of Figure 1 indicates that the stellar enhancement occurred due to the detected stars in the optical bands instead of the detected stars in the IR bands. This stellar enhancement may depend on the values of MSF, whereas MSF depends on the incompleteness of the data as prescribed in Subsection \ref{msf}. In this old cluster (${\bf \sim~1~Gyrs}$), stellar enhancement found due to fainter stars and region may fulfilled by the stellar dust. The IR radiation is easily transmitted from dusty environment, whereas the visual radiation is {\bf obscured} by dust particle or interstellar clouds. The dip in RDPs of the IR range indicates that minimum IR-excesses stars are available in core compare to corona and field region. Thus, large fraction of fainter stars is not detected within the core, which is contradictory {\bf to the} cluster behaviour, whereas RDP of detected stars (minimum in two bands) supported its cluster behaviour. Since, {\bf most} OSCs are too far away therefore too faint to be observable \citep{Zejda+2012}. Similarly, a substantial portion of open clusters is hidden behind interstellar material in the Galaxy plane \citep{Froebrich+2007}. So, this system is considered to be either group of fainter stars or data fluctuation due to PPMXL photometry.\\
The RDPs of optical bands of cluster (depicted in Figure 1) provides the different values of radius, core radius and limiting radius, which are summarized in the Table 2. The absolute distance modulus and colour-excess, $E(R-I)$, of cluster are estimated as $10.612~mag$ and $0.13~mag$, respectively. The best fit theoretical isochrone on $RI$ CMD of NGC 110 yielded a distance of $1.325~kpc$ for the cluster. On the behalf of Figure 7, {\bf we are concluding} that present LF study of the cluster does not support {\bf to study the mass} function and the mass segregation. Since, this stellar system is {\bf containing} those stars which are emitting more energy in the visual bands instead of IR, therefore, the ultraviolet photometric study of this cluster is needed. The mean-proper motion of NGC 110 in {\bf its} RA and DEC directions was estimated as $3.375{\pm}0.195mas/yr$ and $3.493{\pm}0.178mas/yr$, respectively. According to Table 4, the proper motions of cluster and field regions {\bf are} similar in the various radial zones; which indicates that this cluster  is merging with surrounding field region.  We have strongly emphasized that the deep $UBVRI$ photometry of this system is required due to the poor photometric quality of $PPMXL$ catalogue.

\section*{Acknowledgements}
GCJ is thankful to the referee/editor for his vital comments and suggestions which helped us to improve the scientific contexts. GCJ is acknowledge the AP Cyber zone (Nanakmatta) for providing computer facilities, which helped to him for improving the scientific content of this paper. He is also thankful to Dr. R. K. Tyagi (Department of Physics, Govt. P. G. college, Kotdwar, Uttarakhand) for providing the important scientific fact, which are used to develop the context of this paper. This publication makes use of data products from the Two Micron All Sky Survey, which is a joint project of the University of Massachusetts and the Infrared Processing and Analysis Center/California Institute of Technology, funded by the National Aeronautics and Space Administration and the National Science Foundation. This publication makes use of data products from the Wide-field Infrared Survey Explorer, which is a joint project of the University of California, Los Angeles, and the Jet Propulsion Laboratory/California Institute of Technology, funded by the National Aeronautics and Space Administration.

%
\makeatletter
\def\@biblabel#1{}
\makeatother


\received{\it *}
\end{document}